\documentclass[12pt,twoside]{article}

\usepackage{graphics}
\usepackage{epsf}

\newcommand{\Niezurawski}{Nie\.zurawski}
\newcommand{\Zarnecki}{\.Zarnecki}

\def\figheight{0.55\textwidth}

\def\ie{\textit{i.e.\ }}
\def\etal{\textit{et al.}}

\def\bc{\begin{center}}
\def\ec{\end{center}}

\def\bmp{\begin{minipage}}
\def\emp{\end{minipage}}

\def\ar{\rightarrow}
\def\ga{\gamma}
\def\gaga{\ga\ga}

\def\Qbar{\bar{q}}
\def\QQbar{q\Qbar}

\def\bbar{\bar{b}}

\def\bbbar{ b\bbar }

\def\higgsm{\mathit{A/H}}

\def\higgsbb{ \higgsm \ar \bbbar }
\def\higgsgg{ \higgsm \ar g g }

\def\Brhiggsbb{{\rm BR}(\higgsbb)}

\def\gagaQQ{ \gaga \ar \QQbar }
\def\gagaQQg{ \gagaQQ (g) }

\def\gagabb{ \gaga \ar \bbbar }

\def\WW{W^{+} W^{-}}

\def\gagaWW{\gaga \ar \WW}

\def\epem{ e^{+} e^{-} }

\newcommand{\gagahad}{\( \gaga \ar  \mathit{hadrons} \)}

\newcommand{\btagging}{\( b \)-tagging}

\def\AO{A}
\def\HO{H}

\def\MAO{ M_{\AO} }

\def\MAOeq{$ \MAO = $ }

\def\AOHO{ \AO/\HO }
\def\MAOHO{ M_{\AOHO} }

\def\AHbb{ \AOHO \ar \bbbar }

\def\gagaAHbb{ \gaga \ar \AHbb }

\def\tanb{\tan \beta}

\def\tbseven{\tanb = 7}

\def\yJcut{y^{J}_{cut}}
\def\yDcut{y^{D}_{cut}}

\newcommand{\pnfiggeneral}[5]{
\begin{figure}[#1]
{\centering \resizebox*{!}{#2}%
{#3} \par}
 
\caption[]{\label{#4}
#5
}
\end{figure}
}

\newcommand{\pnfig}[5]{
\pnfiggeneral{#1}{#2}{\includegraphics{#3}}{#4}{#5}
}

\def\Pythia{\textsc{Pythia}}

\def\Simdet{\textsc{Simdet}}

\def\Hdecay{\textsc{Hdecay}}

\newcommand{\tgb}{{\rm tg}\beta}
\newcommand{\beq}{\begin{equation}}
\newcommand{\eeq}{\end{equation}}

\begin{document}

\title{Heavy neutral MSSM Higgs Bosons at the PLC \\ - a comparison of two analyses}

\author{M.~Spira$^1$, P.~\Niezurawski$^2$, M.~Krawczyk$^3$, A.~F.~\Zarnecki$^2$ \\[5mm] 
\small\it $^1$ Paul-Scherrer-Institut, CH--5232 Villigen PSI, Switzerland, \\[-2mm]
\small\it $^2$ Institute of Experimental Physics, Warsaw University, ul.\ Ho\.za 69, 00-681 Warsaw, Poland,\\[-2mm]
\small\it $^3$ Institute of Theoretical Physics, Warsaw University, ul.\ Ho\.za 69, 00-681 Warsaw, Poland
}

\date{}

\maketitle

\begin{abstract}

Measurement of the heavy neutral MSSM Higgs bosons $H$ and $A$ production in the process $\gagaAHbb$ 
at the Photon Linear Collider has been considered 
in two independent analyses\cite{MMuhlleitner,MSSM_NZK_PLC2005Proceedings}
for the parameter range corresponding to the so-called "LHC wedge". 
Significantly different conclusions were obtained; signal to background ratio 36 vs.\ 2.  
Here assumptions and results of these two analyses are compared.
We have found that differences in the final results are mainly due to
different assumptions on $\gaga$-luminosity spectra, jet definitions
and selection cuts.

\end{abstract}



\vspace*{4mm}
%
A photon-collider option of the future $\epem$ linear collider 
offers  a unique possibility to produce neutral Higgs bosons 
as $s$-channel resonances.
In this contribution two analyses \cite{MMuhlleitner,MSSM_NZK_PLC2005Proceedings} are compared
which estimate the precision of the cross section measurement 
for the production of heavy neutral MSSM Higgs bosons 
in the process $\gagaAHbb$. 
Both analyses were focused on the so-called ``LHC wedge'', 
\ie the region of intermediate values of $ \tan \beta$, $ \tan \beta \approx$ 4--10,
and masses $\MAOHO$ above 200~GeV,
where the heavy  bosons $\AO$ and $\HO$ may not be discovered 
at the LHC 
and at the first stage of the $\epem$ linear collider. 
In each of these analyses NLO corrections to signal and background processes were taken into account.
As the results of 
the two approaches seem to differ significantly, 
we undertook the task of comparing them,  
focusing on the case of \MAOeq 300~GeV with MSSM parameters $\tgb=7$ 
and $M_2=\mu=200$~GeV. 
%

\begin{figure}[thp]
\vspace*{-2.0cm}
\hspace*{0.2cm}
\epsfxsize=10cm 
\centerline{\epsfbox{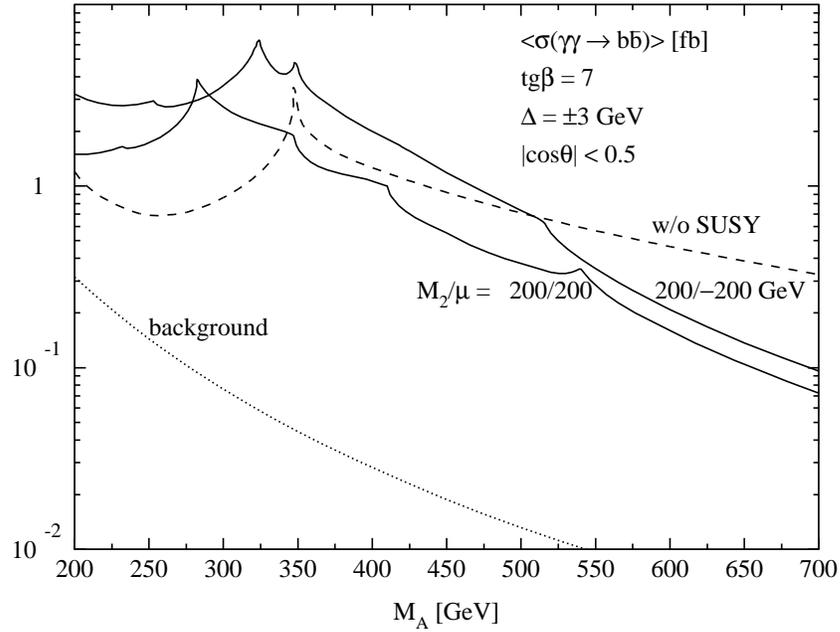}}
\vspace*{-3.5cm}

\caption[]{\it \label{fg:bot} Average cross sections in the invariant mass window $\pm 3$ GeV 
  for resonant heavy Higgs boson
  $H,A$ production in $\gamma\gamma$ collisions as a function of the
pseudoscalar Higgs mass $M_A$
  with final decays into $b\bar b$ pairs, and the corresponding
  background cross section. The maximum of the photon luminosity
  has been tuned to $M_A$.
  The  MSSM parameters have been chosen as $\tgb=7, M_2=\pm \mu=200$ GeV; 
  the limit of vanishing SUSY-particle contributions is shown 
  for comparison. From Ref.~\cite{MMuhlleitner}.}
\end{figure}

In the first analysis \cite{MMuhlleitner} 
the NLO corrections to the background process
$\gamma\gamma \to b\bar{b}$ 
have been calculated according to Ref.~\cite{bkgqcd}. 
Resummation of large Sudakov and non-Sudakov logarithms due to soft gluon 
radiation and soft gluon and bottom-quark 
exchange in the virtual corrections has been taken into account \cite{resum}.
The NLO-$\alpha_s$ was normalized to $\alpha_s(M_Z)=0.119$ 
and the scale given by the $\gaga$ invariant mass was used. 
In order to suppress gluon radiation 
slim two-jet configurations in the final state were selected;
jets were defined within the Sterman--Weinberg criterion. 
If the radiated gluon energy was larger than 10\% of the total $\gamma\gamma$ 
invariant energy and if, at the same time, the
opening angles between all three partons in
the final state were larger than $20^o$, the event was classified as three-jet event and rejected.
The interference between the signal and background processes has
been taken into account properly.  
The NLO QCD corrections of the interference terms to quark
final states including the resummation of the large (non-)Sudakov logarithms
were calculated (for details see \cite{muehldiss}). 
The description of the $\gamma\gamma$-luminosity was based on the LO cross
section formula for the Compton process.
It was assumed that  the beam energy is tuned to obtain maximum luminosity at the value
of the pseudoscalar Higgs mass $M_A$.
In this analysis the selection of events was applied at the parton level.
The background was reduced with a cut in the production angle of the bottom quark only, $|\cos\theta_b | < 0.5$.
By collecting $b\bar b$ final states with a resolution in
the invariant mass $M_A \pm 3$ GeV, the sensitivity to the
combined $A$ and $H$ resonance peaks above the background was strongly
increased.
The result for the peak cross section is shown in Fig.~\ref{fg:bot} as
a function of~$M_A$. 
It can be inferred from the figure that the background is strongly
suppressed (signal to background ratio $\approx 36$ for $M_A=300$ GeV).
%

%
The second analysis \cite{MSSM_NZK_PLC2005Proceedings}
was based on realistic simulations of the $\gaga$-luminosity spectra 
for the PLC at the TESLA collider \cite{V.TelnovSpectra,CompAZ}. 
One-year run of PLC was assumed with the center-of-mass energy of the
colliding electron beams
optimized for the production of the pseudoscalar Higgs bosons.
The distribution of the primary vertex 
and the beam crossing angle were taken into account.
The total widths and branching ratios of the Higgs bosons and the $\HO$ mass 
were calculated with the program \Hdecay{} \cite{HDECAY},
taking into account decays to and loops of supersymmetric particles.\footnote{  The MSSM parameter set considered here was denoted in \cite{MSSM_NZK_PLC2005Proceedings} as $I$.}
These results were used to calculate the signal cross section in the
resonance approximation. The events were generated
with the \Pythia{} program \cite{PYTHIA}.
As the main background to Higgs-boson production 
heavy-quark pair production was considered;
the event samples were  generated using 
the program by G.~Jikia \cite{JikiaAndSoldner} 
which includes exact one-loop QCD corrections to the lowest order processes $\gagaQQg$ \cite{JikiaAndTkabladze}, 
and the non-Sudakov form factor in the double-logarithmic approximation, 
calculated up to 4 loops \cite{MellesStirlingKhoze}.
The JADE jet definition with a parameter
$\yJcut=0.01$ is used to define 2- and 3-parton final states.
The resummation of Sudakov logarithms due to soft gluon bremsstrahlung  is omitted. 
In the calculation the LO-$\alpha_s$ normalized to $\alpha_s(M_Z)=0.119$ was used
at the scale given by the average of the squared transverse masses of the quark and anti-quark. 
Other background processes 
were also studied.
As about  two  \gagahad{} events (so-called {\em overlaying events})
are expected   per bunch crossing,
they were generated with \Pythia{},
and have been overlaid on signal and background events
according to the Poisson distribution.  
On the detector level\footnote{
The detector performance was simulated
by the program   \Simdet{}  \cite{SIMDET401}.
}
jets were reconstructed using the Durham algorithm with $\yDcut=0.02$.
Two or three jet events were accepted.
To reduce the heavy-quark production background 
a cut on the polar angle for each jet was imposed:   $|\cos {\theta}_{jet}| < 0.65$.
Since the Higgs bosons are  produced almost at rest, 
the ratio of the total longitudinal momentum calculated from all  jets in the event, $P_{z}$,
to the total measured energy, $E$, was required to be small: $|P_{z}|/E < 0.06$.
Additional cuts to suppress the influence of overlaying events and the $\gagaWW$ background were also applied.
A realistic \btagging{} algorithm was  used.
All cuts were optimized.
A more detailed description of event generation, simulation and selection cuts
can be found in \cite{SM_PN,MSSM_PN}.
The result of the analysis  is shown in Fig.\ \ref{fig:plot_var34_m300_modmssm_oe1_costhtc0.85} 
where the distribution of the corrected invariant mass, $W_{corr}$ 
(correction for escaping neutrinos; see \cite{NZKhbbm120appb}), 
after imposing all selection cuts is presented for the signal and individual background contributions.
From the number of signal and background events in 
the optimized $W_{corr}$-window the expected statistical precision
of the cross-section measurement for \MAOeq 300 GeV is equal to 11\%.

The results of both analyses differ significantly. 
In the first analysis the background contribution is negligible: 
the signal to background ratio is $S/B \approx 36$ in the invariant mass window 297-303 GeV. 
For comparison in the second analysis $S/B \approx 2$ was obtained 
in the corrected invariant mass window 295-305 GeV 
if only the process $\gagabb$ is taken into account as the background. 
%
In order to understand the sources of those
differences the cross sections for the background process $\gagabb$ and signal process $\gagaAHbb$ 
were recalculated within both approaches
with the same cuts and the same $\gaga$-luminosity spectrum.
The following conclusions emerged after investigation of 
the two calculations of the heavy quark background.
With the polar angle cut imposed only on the quark $b$
the 3-jet part is larger than the 2-jet part by more than an order of magnitude. 
However, if the cut on the anti-quark angle is added, 
the 2-jet and 3-jet cross sections differ only by a factor 2 to 3.
Thus, requiring only 2-jet events is less essential 
if the angular cut is applied for both quarks.
This corresponds to the common cut on the jet polar angle
which is usually applied on the detector level.
The 2-jet cross sections obtained in the two approaches agree
within a factor of 2.
Moreover, the full resummation of Sudakov and non-Sudakov logarithms
does not modify the 2-jet numbers too much compared to the 4-loop
expansion of the non-Sudakov logarithms.
If the JADE algorithm is applied in both analyses 
then the obtained cross sections agree within 15\%.
These residual differences can be mainly attributed to the different scale
choices of the strong coupling $\alpha_s$.
Next, the analogous comparison for the signal process was performed for \MAOeq 300~GeV.
The same MSSM parameter set was used, \ie $ \tbseven , \, \mu = 200 $~GeV, $M_{2} = $ 200~GeV,   
trilinear couplings equal to 1500~GeV, and common sfermion mass equal to 1~TeV were assumed.
A top quark mass of 174~GeV has been adopted.
Decays to supersymmetric particles and loops with them  were taken into account. 
Small contributions from the decays $\higgsgg^* \ar g \bbbar$ were not added to the branching ratio $\Brhiggsbb$.
With JADE jet definitions the results of both approaches agree within
5\% for the total cross section, 
and within 30\% for the 2-jet and 3-jet classes separately.
The differences in the separation of 2-jet and 3-jet classes mainly
originate from the different approaches used in the two analyses. The
second analysis used the resonance approximation and generated gluon
radiation by parton showers, while the first analysis used a full NLO
calculation for the signal process including soft gluon resummation for
the 2-jet part.
Finally, we have compared the results for the invariant-mass window
297-303 GeV taking into account the assumed $\gaga$-luminosity spectra with the same normalization.
Our first conclusion is that if the JADE jet definition were used in
both analyses, the difference in the signal to background ratio between
our analyses would be mainly due to the different contributions
of $J_z=0$ and $|J_z|=2$ parts to the $\gaga$-luminosity.
The $J_z=0$ luminosity component of the realistic luminosity distribution used in the second analysis 
amounts only to 94\% of the same component of the ideal spectrum used in
the first analysis. 
What is more important, in the realistic spectrum about 5.5 times more of the
$|J_z|=2$ component is taken into account relative to the same component in the ideal spectrum.
If the JADE algorithm  with $\yJcut=0.01$  is used,  the signal to
background ratio is around 12 in case of the first approach with angular
cuts $|\cos\theta_{b/\bar{b}} | < 0.5$ applied for each quark and if 2-
and 3-jet events are taken into account. 
In the second approach the ratio is around 6.
However, if a correction accounting for the differences in the
luminosity spectra is applied, the rescaled result of the second
analysis is around 10, thus only 17\% less than in the first analysis.
Our second observation is that the use of the Sterman-Weinberg jet
definition leads to much higher rates of 2-jet events for the signal
than for the background. 
This results in nearly 2 times higher signal to background ratios 
in comparison to results obtained with the JADE jet definition if only 2-jet events are taken into account.
The measurement of the process $\gagaAHbb$ at the PLC is very promising,
even for the realistic $\gaga$-luminosity spectrum, which is
less advantageous  than the ideal one. 
Use of the clustering algorithm based on the Sterman-Weinberg jet
definition would lead to much higher signal to background ratios, if
only 2-jet events were taken into account. 

M.~K., A.~F.~\.Z., and P.~N.~acknowledge partial support by Polish Ministry of Science and Higher Education, 
grant no.\ 1 P03B 040 26 and M.~S.~partial support by the Swiss
Bundesamt f\"ur Bildung und Wissenschaft.

\pnfig{bht}{\figheight}{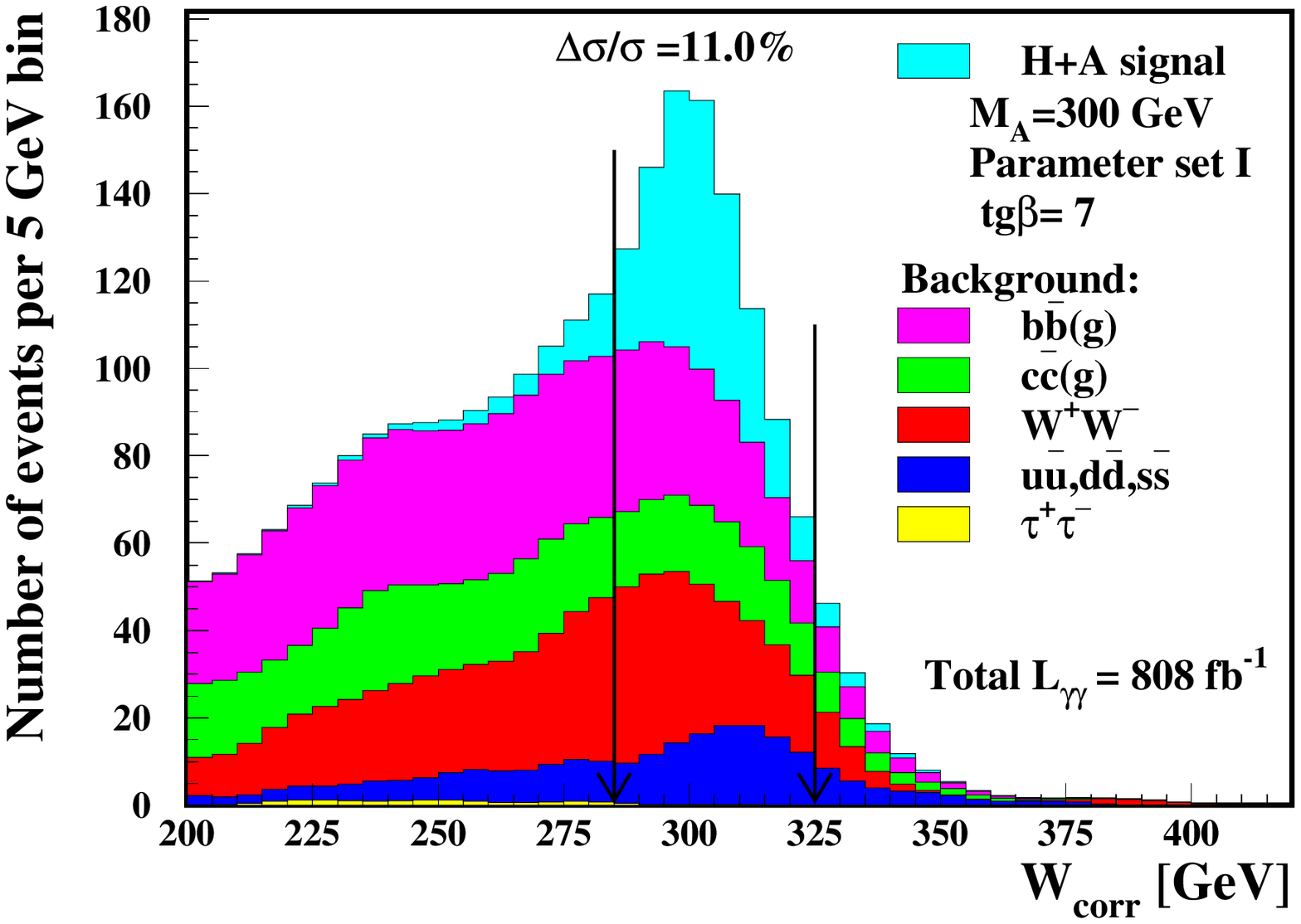}{fig:plot_var34_m300_modmssm_oe1_costhtc0.85}{
\it Distributions of the corrected invariant mass, $W_{corr}$, for
signal and all considered background contributions, with  overlaying
events included.
The best precision of 11\% for $\gagaAHbb$ cross section measurement is
achieved in the $W_{corr}$ window between   285 and  325~GeV. From
Ref.~\cite{MSSM_NZK_PLC2005Proceedings}.
}

\end{document}